\documentclass[aps,pra,preprint,amsmath,amssymb,nofootinbib,superscriptaddress]{revtex4}

\newcommand{\ket}[1]{|#1\rangle}

\usepackage[pdftex]{graphicx}
\usepackage{mathrsfs}
\usepackage{color}
\usepackage[colorlinks]{hyperref}

\begin{document}

\bibliographystyle{nature}


\title{Self-avoiding quantum walks}


\author{Elizabeth Camilleri}
\affiliation{Centre for Engineered Quantum Systems, Department of Physics and Astronomy, Macquarie University, Sydney NSW 2113, Australia}

\author{Peter P. Rohde}
\affiliation{Centre for Engineered Quantum Systems, Department of Physics and Astronomy, Macquarie University, Sydney NSW 2113, Australia}

\author{Jason Twamley}
\affiliation{Centre for Engineered Quantum Systems, Department of Physics and Astronomy, Macquarie University, Sydney NSW 2113, Australia}

\date{\today}

\frenchspacing


\begin{abstract}
Quantum walks exhibit many unique characteristics compared to classical random walks. In the classical setting, self-avoiding random walks have been studied as a variation on the usual classical random walk. Classical self-avoiding random walks have found numerous  applications, most notably in the modeling of protein folding. We consider the analogous problem in the quantum setting. We complement a quantum walk with a memory register that records where the walker has previously resided. The walker is then able to avoid returning back to previously visited sites. We parameterise the strength of the memory recording and the strength of the memory back-action on the walker's motion, and investigate their effect on the dynamics of the walk. We find that by manipulating these parameters the walk can be made to reproduce ideal quantum or classical random walk statistics, or a plethora of more elaborate diffusive phenomena. In some parameter regimes we observe a close correspondence between classical self-avoiding random walks and the quantum self-avoiding walk.
\end{abstract}

\maketitle



Quantum walks \cite{bib:ADZ, bib:AAKV, bib:Kempe08, bib:Salvador12}, the quantum equivalent of classical random walks, have been studied extensively for their applications in quantum information processing \cite{bib:NielsenChuang00}. Here a \emph{walker} (a particle such as a photon) resides at a vertex (e.g. an optical mode) in a graph and is allowed to `hop' along the edges in the graph to reach other vertices. In the classical random walk this process takes place randomly, whereas in the quantum case the walker coherently enters a superposition across different vertices. Numerous elementary optical demonstrations of quantum walks have been performed \cite{bib:Hagai08, bib:Schreiber10, bib:Broome10, bib:Peruzzo10, bib:Schreiber11b, bib:Matthews11, bib:Owens11, bib:Schreiber12, Sansoni12}, experimentally confirming the unique behaviour of quantum walks compared to classical random walks, and providing an alternate route towards optical quantum information processing \cite{bib:KLM01, bib:KokLovett11}.

In this paper we consider quantum walks, where the walker has memory of its previous location history and the coin operator is a function of the memory. This can be used, for example, to preferentially avoid previously visited vertices or implement more elaborate conditional coins. Self-avoiding walks have been studied extensively in the classical case \cite{bib:Amit83, bib:Byrnes84, bib:Schulz05, bib:Toth10, bib:Slade11, bib:MadrasSlade13}, having been applied to applications such as protein folding and percolation theory.

Previous authors have considered quantum walks complemented by memory, where the memory is of previous coin values \cite{bib:Brun03b, bib:Flitney04, bib:McGettrick10, bib:RohdeBrennen12}. Here, however, we consider the case where the memory is of previous positions rather than of previous coins.

Self-avoiding quantum walks may find applications in situations where ordinary quantum walks have been applied, but where we explicitly wish to avoid re-exploring previously visited sites. While classical self-avoiding walks have been applied to modelling protein folding \cite{bib:Bahi13}, long-chain polymers \cite{bib:Flory49} and molecular conformation, perhaps quantum self-avoiding walks may be applied to modelling the analogous problems in the quantum context.

In our model each vertex in the graph is complimented by a qubit, which marks whether the site has previously been visited. As the walker walks it leaves a record on the site at which it is presently located. The coin, which determines the subsequent step direction, is biased to avoid walking onto previously marked sites. We will see that due to the unitarity of our model, we cannot maintain a permanent record of where the walker has been and the walker has some chance of revisiting previously visited sites.


We find that self-avoiding quantum walks exhibit diverse diffusive characteristics, ranging from ideal classical to ideal quantum walk dynamics, and a plethora of more elaborate behaviours. In the limit of maximum self-avoidance, the walk can exhibit significantly faster rate of spread than an ordinary quantum walk.


\section*{Classical self-avoiding random walks} 

Self-avoiding classical random walks \cite{bib:Amit83, bib:Byrnes84, bib:Schulz05, bib:Toth10} are a variation on the standard classical walk, where the walker is unable to revisit any sites it has previously visited. A one dimensional, classical, completely self-avoiding walk is trivial; once the first step has been taken by the walker, every step thereafter has only one possibility. The variance of the walker will always be zero, as after the initial step there is only one possible location for the walker to be at any given step. Self-avoiding random walks become less trivial and more useful in two and three dimensions, but they also become exceedingly difficult to analyse in higher dimensions.

Self-avoiding walks have the property that they explore a graph more rapidly, since evolution time is not wasted re-exploring already visited sites. Self-avoiding classical random walks are particularly useful in two and three dimensions as they find applications in modelling solvents and polymers \cite{bib:Flory49}, protein folding \cite{bib:Bahi13}, molecular conformation, percolation lattices, and drug discovery. When using classical random walks to model the proliferation of forest fires or the propagation of liquids through a porous material -- two archetypal applications for classical random walks -- we wish to determine whether a route through a graph exists. Thus, there is no benefit in exploring vertices that have previously been explored and a self-avoiding walk will be far more efficient.


\section*{Quantum walks} 
A (discrete-time) quantum walk is a bipartite system comprising \emph{position} and \emph{coin} degrees of freedom, with Hilbert space \mbox{$\mathcal{H}=\mathcal{H}_P\otimes \mathcal{H}_C$}. The position parameter specifies the location (vertex) of the walker in the graph, whilst the coin parameter specifies the direction the walker is heading. On a one-dimensional graph the state of the quantum walker at time $t$ is of the form $\ket{\psi(t)} = \sum_{x,c} \alpha_{x,c}(t) \ket{x,c}$, where \mbox{$x \in \mathbb{Z}$} is the discrete position of the walker, and \mbox{$c=\pm 1$} is the direction of the walker (left or right respectively).

The evolution of the walk is specified by unitary \emph{coin} ($\hat{C}$) and \emph{step} ($\hat{S}$) operators, which have the actions $\ket{x,c} \mapsto \sum_{j=\pm1} U_{c,j} \ket{x,j}$ and $\ket{x,c} \mapsto \ket{x+c,c}$ respectively, where $U$ is the unitary coin matrix. Thus the coin operator coherently manipulates the direction of the walker, leaving the position unchanged, whilst the step operator uses the coin parameter to update the location of the walker, leaving the coin unchanged. The evolution of the walk over $t$ time steps proceeds as $\ket{\psi(t)} = (\hat{S}\cdot \hat{C})^t \ket{\psi(0)}$.

One of the interesting features of the quantum walk is that it exhibits a quadratically faster rate of spread (measured using variance) compared to the classical random walk. The variance is defined as, $\sigma^2(t) = \sum_x p_x(t) \cdot (x-\mu(t))^2$, $\mu(t) = \sum_x p_x(t) \cdot x$, where $p_x(t)$ is the total probability of finding the walker at location $x$ at time $t$, given by summing over the coin degree of freedom, $p_x(t) = \sum_c \left| \alpha_{x,c}(t) \right|^2$, effectively tracing out the coin.


\section*{Quantum self-avoiding walks} 
\begin{figure}[!htb]
\includegraphics[width=\columnwidth]{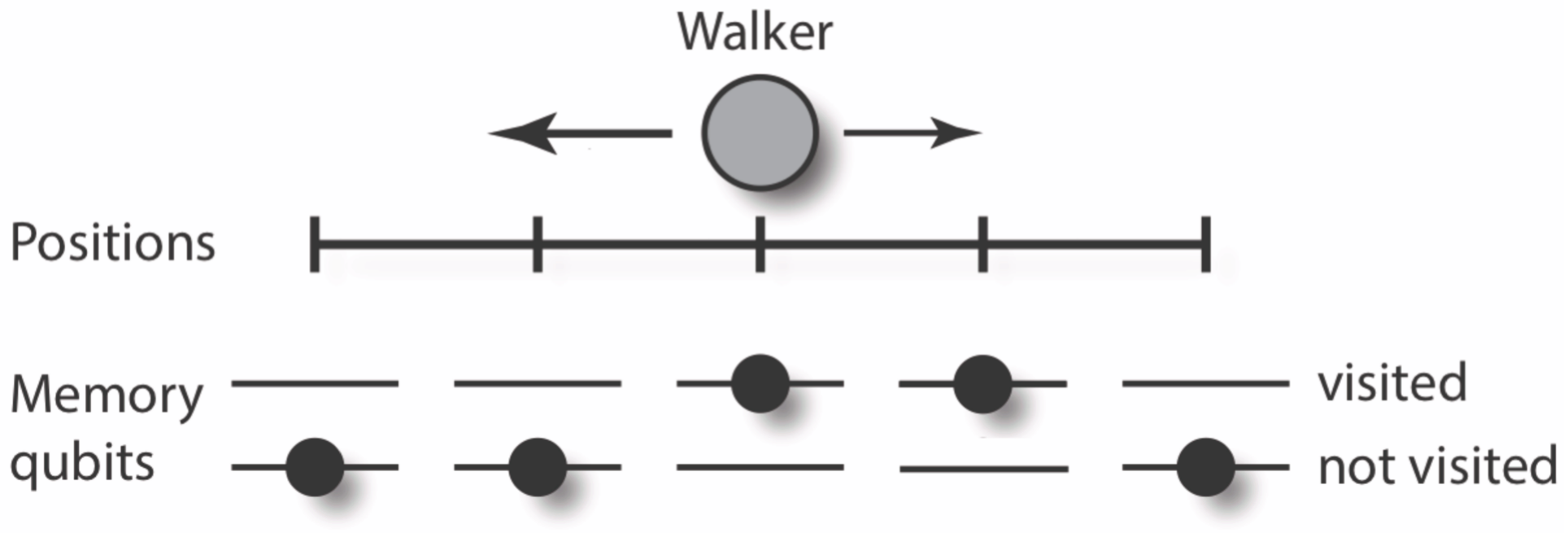}
\caption{Illustration of the self-avoiding quantum walk step: The walker resides on a linear graph structure with a two-level memory qubit associated with each vertex, indicating whether the site has been previously visited. The quantum self-avoiding walk step operates in three parts: (i) the walker records their current position by flipping their local memory qubit, (ii) sets the bias of the coin depending on the neighboring memory sites  to yield motion preferentially onto adjacent sites whose memory qubits are not set, and (iii) conditional move of the walker's position depending on the coin state. The overall action of these combined parts bias the motion of the walker to move preferentially away from sites on the line where it has previously visited. The example configuration depicted in the diagram results in a quantum walk step with an increased amplitude for the walker to move to the left due to the state of the adjacent memory registers.} \label{fig:walker_diagram}
\end{figure}

We fashion our self-avoiding quantum walk with three components: (1) we record the current position of the walker into the local quantum memory. This is performed using a `memory update operator'; (2) we perform the usual coin operator, where the bias of the coin is determined by the state of the memory registers neighbouring the current location of the walker; (3) the usual step operator is applied, which results in the walker moving preferentially onto a neighbour whose memory register indicates that the site has not been visited. An illustration of the components in the self-avoiding quantum walk are shown in Fig.~\ref{fig:walker_diagram}.

\emph{Recording the current position} --- To model the self-avoiding quantum walk whereby the walker has memory of its previous location history and avoids positions where it has previously resided, we complement the walker by a sequence of qubits -- one for each site in the lattice -- where the qubits indicate whether the walker has previously visited the respective site. In this case the state of the system at time $t$ is of the form,
\begin{equation}
\ket{\psi(t)} = \sum_{x,c,q_1,\dots,q_N} \alpha_{x,c,q_1,\dots,q_N}(t) \ket{x,c,q_1,\dots,q_N},
\end{equation}
where \mbox{$q_i=\{0,1\}$} is the qubit associated with site $i$, and there are $N$ sites. Initially \mbox{$q_i=0\,\,\forall\,\, i$}, indicating that none of the sites have been visited. Evidently, the size of the Hilbert space grows exponentially with the size of the lattice (\mbox{$|\mathcal{H}|=N\cdot 2^{N+1}$}), in contrast to linear growth for an ordinary quantum walk (\mbox{$|\mathcal{H}| = 2N$}). This limits our numerical investigations to relatively small evolution times.

We introduce a memory update operator, which updates the memory qubits as a function of the walker's position. We have chosen to implement the memory update operator unitarily, which implies that the memory is reversible and temporary, and that the record will evaporate after a second visit to the respective site. A permanent memory recording would require a Lindblad model, which would introduce decoherence into the system. Taking our memory recording to be unitary we fix it to be of the form,
\begin{equation}
\hat{M}(\theta_M) \ket{x,c,q_1,\dots,q_N} \mapsto \hat{R}^{(x)}(\theta_M) \ket{x,c,q_1,\dots,q_N},
\end{equation}
a Pauli $X$ rotation applied to the $x$th qubit. Now $\theta_M$ determines the strength of the action of the position on the memory. In the extreme case where \mbox{$\theta_M=\pi/2$}, if the walker has not previously visited site $x$, the memory update operator flips the respective qubit to mark that this position has now been visited. Of course, a second visit to the site will flip the qubit again, marking that the site has not been visited. Note that this latter property is unavoidable if the evolution is to remain unitary. If \mbox{$\theta_M=0$} the position has no effect on the memory and the memory only couples with position via the coin operator.

\emph{Biasing the coin according to neighbouring memory registers} --- The coin operator is defined analogously to the usual quantum walk coin operator,
\begin{equation}
\hat{C}(\theta_C,\theta_B) \ket{x,c,q_1,\dots,q_N} \mapsto \sum_{j=\pm 1} U_{c,j}^{(q_x)} \ket{x,j,q_1,\dots,q_N},
\end{equation}
where \mbox{$U^{(q_x)}$} is a \mbox{$2\times 2$} coin matrix that depends on the state of the qubit at location $x$, $q_x$. This operator coherently manipulates the coin degree of freedom locally, whilst leaving the position and memory qubits unchanged. The choice of coin matrix is a function of the local memory qubit -- when \mbox{$q_x=0$} we let \mbox{$U^{(q_x)}=R(\theta_C)=R(\pi/4)$} (a usual balanced walk with equal amplitudes of stepping in either direction), and when \mbox{$q_x=1$} we let \mbox{$U^{(q_x)}=R(\theta_B)$}. Here, $R(\theta)$ is a general Pauli $X$ rotation,
\begin{equation}
R(\theta) = 
\left(\begin{array}{c c}
\mathrm{cos}\,\theta & -i \,\mathrm{sin}\,\theta \\
-i \,\mathrm{sin}\,\theta & \mathrm{cos}\,\theta \end{array}\right).
\end{equation}
Thus, \mbox{$\theta=0$} implements an identity gate, \mbox{$\theta=\pi/2$} implements a full bit-flip, and \mbox{$\theta=\pi/4$} implements a balanced step. Our memory will have the effect that if the $x$th qubit is set (indicating that we previously visited location $x$) then the coin operator implements a partially reflecting mirror on the walker's dynamics, with strength determined by $\theta_B$. This mirror is located at spatial location $x$ and the reflectivity ($r$) is determined by $\theta_B$ (when $\theta_B=0$, $r=0$, and when $\theta_B=\pi/2$, $r=1$). Thus, the parameter $\theta_B$ controls how strong the effect of memory on the coin is. The probability of being reflected at a site where the memory is set is given by \mbox{$r=|\mathrm{sin}\,\theta_B|^2$}, otherwise it is given by \mbox{$r=|\mathrm{sin}\,\theta_C|^2$}.

\emph{Propagating the walker} --- The step operator is defined in the usual way -- it updates the position as a function of the coin parameter,
\begin{equation}
\hat{S} \ket{x,c,q_1,\dots,q_N} \mapsto \ket{x+c,c,q_1,\dots,q_N}.
\end{equation}
The step operator affects only the position, leaving the coin and memory systems unchanged.

\emph{Combined evolution} --- The total evolution now progresses as,
\begin{equation}
\ket{\psi(t)} = [\hat{S}\cdot \hat{C}(\theta_C,\theta_B) \cdot \hat{M}(\theta_M)]^t \ket{\psi(0)},
\end{equation}
where a single step begins with memory recording, followed by the coin operator, and finally the step operator.

Our formalism differs from that presented in Ref. \cite{bib:BarrProctor13}, where no memory was required to ensure self-avoidance. Rather, the coin matrix was chosen so as to guarantee there is always zero amplitude in the reverse direction. This is the so-called `non-reversing quantum walk'. However, such a formalism is constrained by the fact that one is unable to tune between different memory strengths, making our formalism much more versatile, as we will see in the next section.


\section*{Results}
\begin{figure*}[!htb]
\includegraphics[width=1\columnwidth]{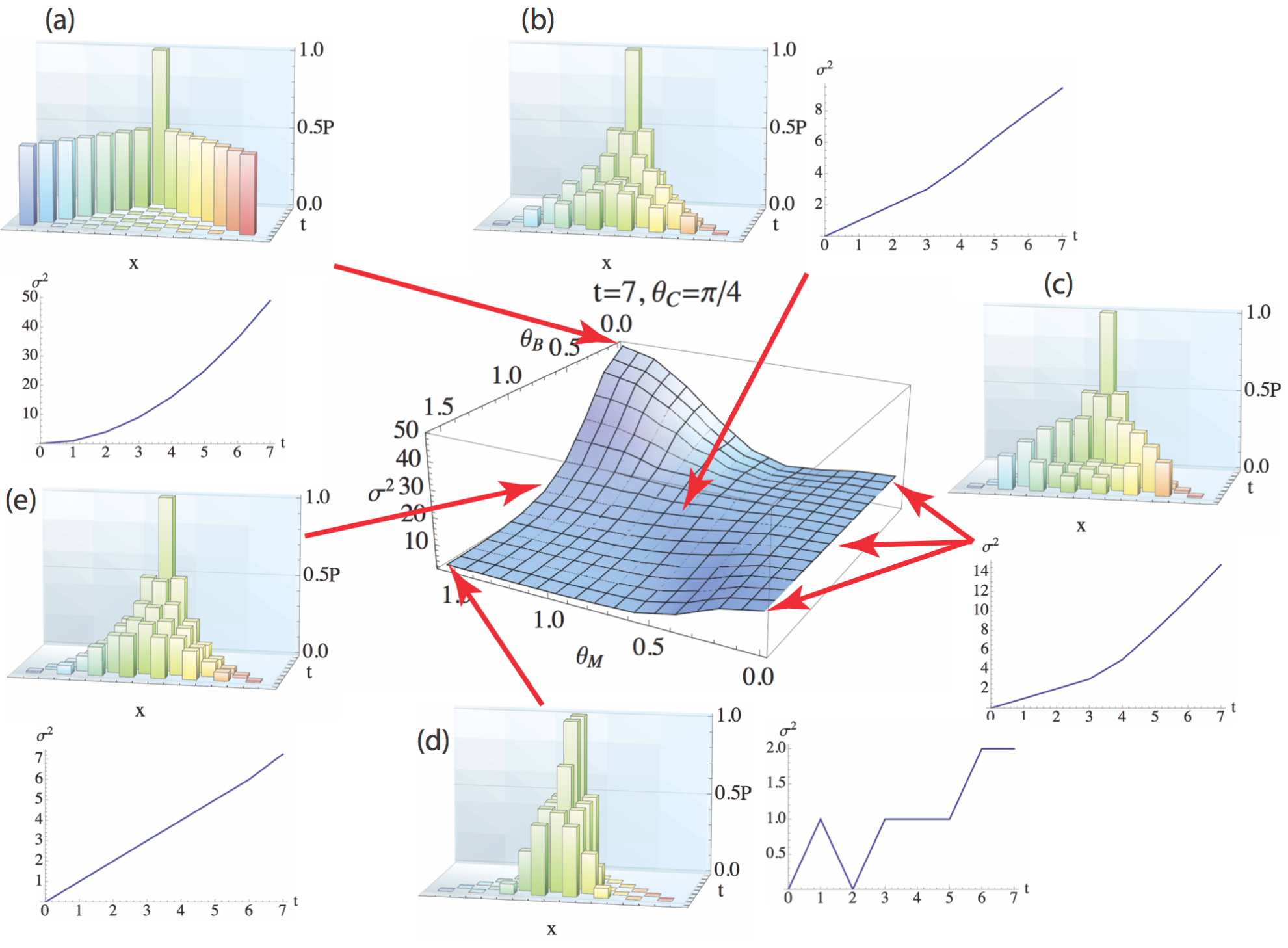}
\caption{(Colour online) Final variance ($t=7$) of the self-avoiding quantum walk with symmetrised input state (Eq.~\ref{eq:input_sym}), against memory recording strength ($\theta_M$) and coin back-action strength ($\theta_B$), and, as insets, some of the corresponding quantum walks (showing the full time evolution of the variance and probability distribution). (e) In the limit of $\theta_M=\pi/2$ and $\theta_B=\pi/4$ we observe ideal classical random walk statistics, since the memory is maximally entangled with the walker, yielding a balanced $\pi/4$ walk with decoherence upon tracing out the memory subsystem. (c) In the opposing limit of $\theta_M=0$ and irrespective of $\theta_B$ we observe ideal quantum walk statistics, since now the memory does not couple with the walker, yielding a uniform $\pi/4$ walk without decoherence. The memory never changes, remaining at its initial value. (a) Variance is maximised when $\theta_M=\pi/2$ and $\theta_B=0$, for which the coin will always be the identity coin, yielding two outward straight line trajectories. (d) When $\theta_M=\pi/2$ and $\theta_B=\pi/2$ the dispersion is restricted since the walker is switching between a balanced $\pi/4$ coin and a bit-flip coin at alternate steps.} \label{fig:composed_symmetric}
\end{figure*}

To characterise the self-avoiding walk we will again consider the variance of the position probability distribution against time. We compute the variance, where now $p_x(t)$ is obtained by summing over the coin and memory degrees of freedom,
\begin{equation}
p_x(t) = \sum_c \sum_{q_1,\dots,q_N} \left| \alpha_{x,c,q_1,\dots,q_N}(t) \right|^2,
\end{equation}
effectively tracing out the coin and memory subsystems.

In the purely classical case $\sigma^2(t)\propto t$, and in the quantum case $\sigma^2(t)\propto t^2$. In our quantum walk with memory we anticipate more elaborate dynamics, which we will now explore and characterise.

In \mbox{Fig.~\ref{fig:composed_symmetric}(centre)} we plot the variance of the self-avoiding quantum walk against varying memory recording ($\theta_M$) and coin back-action ($\theta_B$) when evolved 7 steps. In all cases the regular coin is fixed as a balanced coin, $\theta_C=\pi/4$. Here we have employed the symmetrised input state,
\begin{equation} \label{eq:input_sym}
\ket{\psi(0)} = \ket{1_x} \otimes \frac{1}{\sqrt{2}} (\ket{1_c} + \ket{-1_c}) \otimes \ket{0_{q_1},\dots,0_{q_N}},
\end{equation}
which guarantees symmetric position probability distributions in the case of the standard unbiased quantum walk.

Each point in the plot in \mbox{Fig.~\ref{fig:composed_symmetric}(centre)} corresponds to the final variance of an entire quantum walk with given choices of coin back-action and memory recording. To understand these dynamics, the insets illustrate the complete quantum walk evolution dynamics corresponding to some individual points in the plot.

We observe that the variance is maximal when the memory has maximum recording strength (\mbox{$\theta_M=\pi/2$}) and the coin has identity back-action (\mbox{$\theta_B=0$}), as shown in \mbox{Fig.~\ref{fig:composed_symmetric}(a)}. Because the memory recording strength is maximal, the memory qubit corresponding to the initial location is set. Then, because the qubit is set, the identity coin will be applied, which propagates the walker outwards. At the next step the same process applies, marking the current site and propagating the walker outwards. Thus, the two terms in the symmetrised input state shoot off in straight line paths in opposite directions, yielding maximum quadratic variance against time. Fitting to a second order polynomial, 
\begin{equation}
\sigma^2(t) = k_0 + k_1 t + k_2 t^2,
\end{equation}
the second order coefficient of the aforementioned maximum variance walk is \mbox{$k_2=1$}, compared to \mbox{$k_2=0.27$} for the ideal quantum walk, reflecting the significantly faster rate of spread of the \mbox{$\theta_M=\pi/2$}, \mbox{$\theta_B=0$} self-avoiding quantum walk compared to the normal ideal quantum walk.

\begin{figure}[!htb]
\includegraphics[width=0.48\columnwidth]{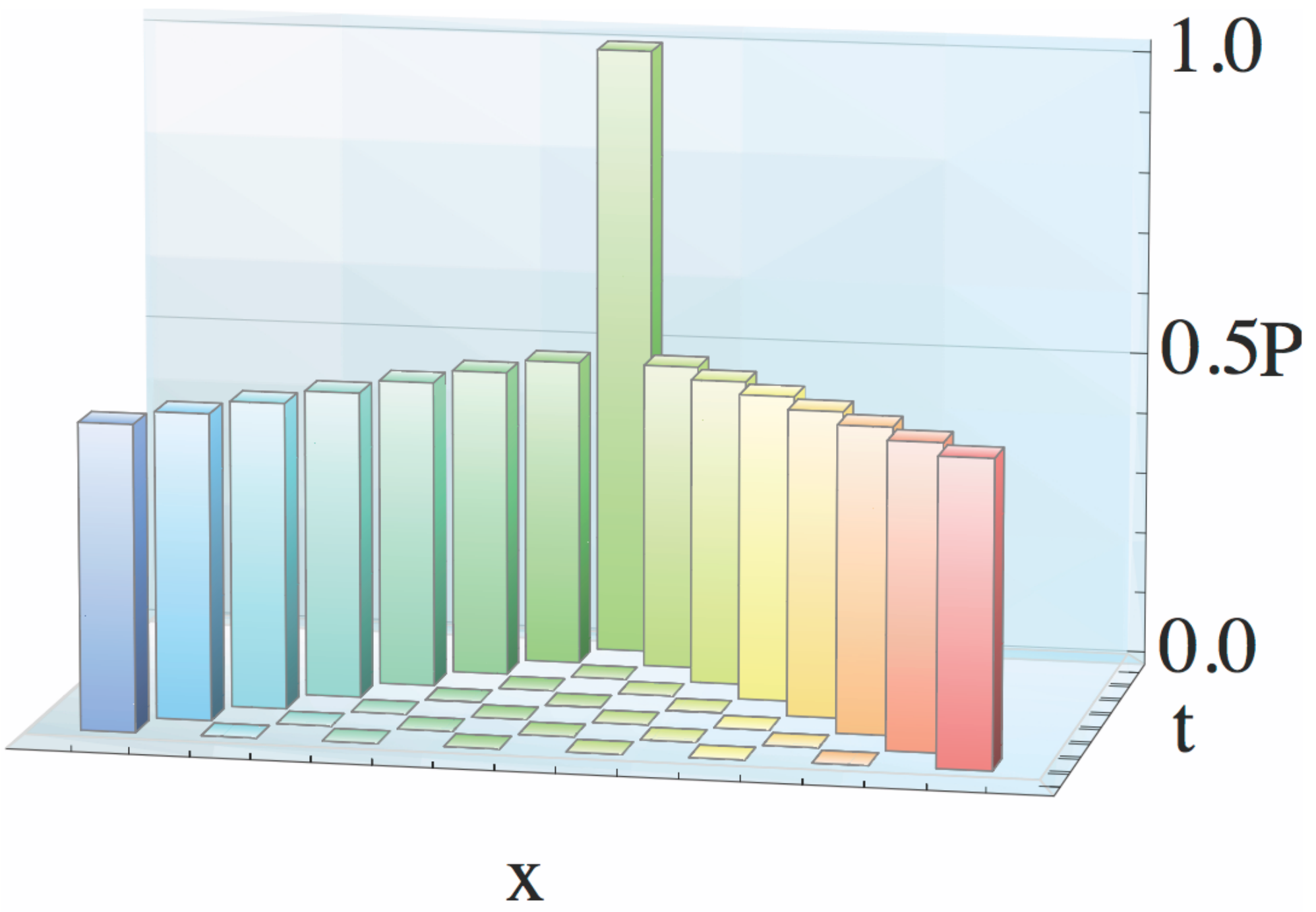}
\includegraphics[width=0.48\columnwidth]{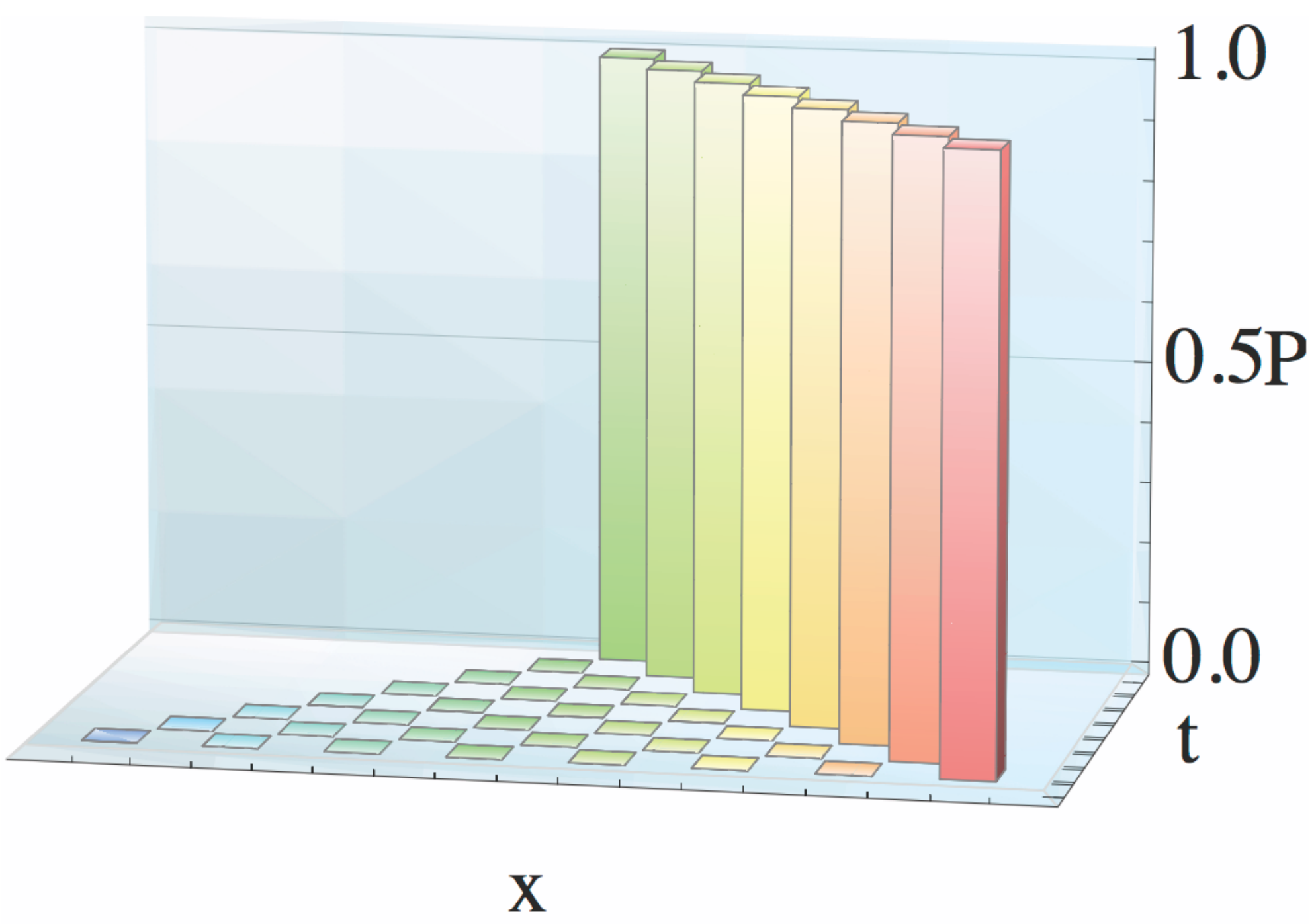}
\caption{(Colour online) Evolution of the self-avoiding quantum walk for $\theta_M=\pi/2$, $\theta_B=0$, with symmetrised (left) and unsymmetrised (right) input states (Eqs.~\ref{eq:input_sym} and \ref{eq:input_asym} respectively). This illustrates the high sensitivity of the dispersion to the initial state, as the former exhibits the maximum possible variance, whereas the latter exhibits zero variance.} \label{fig:max_min_var}
\end{figure}

\begin{figure}[!htb]
\includegraphics[width=0.7\columnwidth]{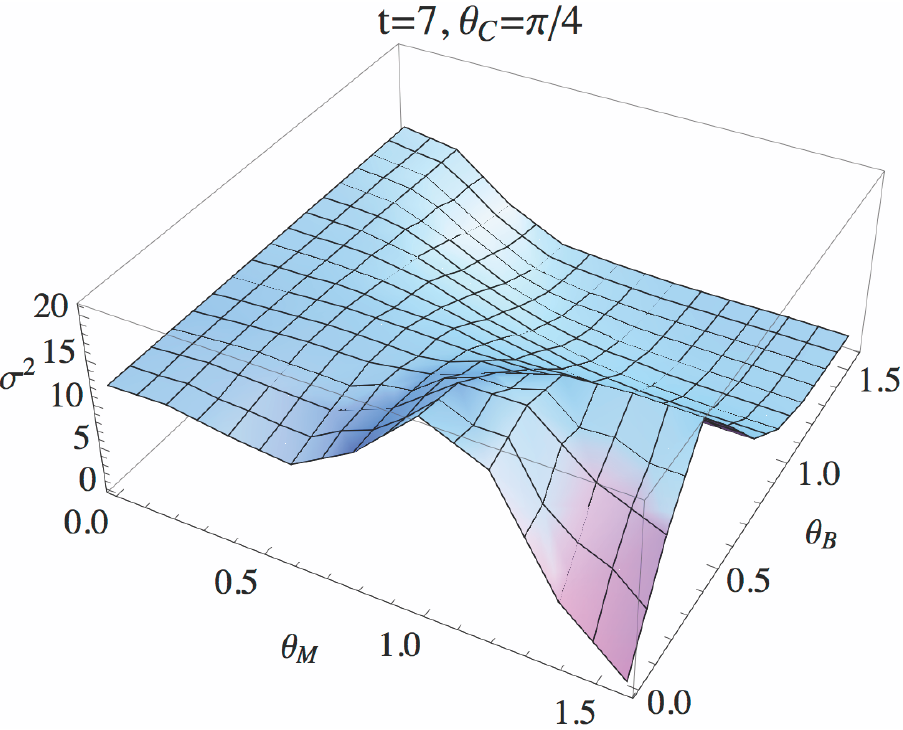}
\caption{(Colour online) Variance of the self-avoiding quantum walk with unsymmetrised input state (Eq.~\ref{eq:input_asym}). Contrast the different dynamics with \mbox{Fig.~\ref{fig:composed_symmetric}} where a symmetrised input state was employed. In particular, note the drastically different variance at \mbox{$\theta_M=\pi/2$}, \mbox{$\theta_B=0$} compared to the symmetrised case -- the maxima in the symmetrised plot corresponds to the minima of the unsymmetrised plot. Note the different viewing angle than \mbox{Fig.~\ref{fig:composed_symmetric}}.} \label{fig:3d_asym}
\end{figure}

Intuitively, the maximum variance in this case appears to be a result of the choice of symmetrised input state. We expect that if we instead adopt an unsymmetrised input state, 
\begin{equation} \label{eq:input_asym}
\ket{\psi(0)} = \ket{1_x,1_c,0_{q_1},\dots,0_{q_{N}}},
\end{equation}
there will be only a single straight line trajectory in one direction, yielding zero variance. This is confirmed in \mbox{Fig.~\ref{fig:max_min_var}}. Thus, we see that the variance of the self-avoiding quantum walk is highly dependent on the initial choice of input state, and can in fact swap us between maximal and zero variance. \mbox{Fig.~\ref{fig:3d_asym}} illustrates the variance of an unsymmeterised input state across all parameter regimes.

In the intermediate case of $0<\theta_B<\pi/2$, the coin creates a superposition, yielding lower variance than \mbox{Fig.~\ref{fig:composed_symmetric}(a)}.

In the case of \mbox{$\theta_M=\pi/4$} and \mbox{$\theta_B=\pi/4$} we observe statistics that are qualitatively very classical (\mbox{Fig.~\ref{fig:composed_symmetric}(b)}). This is because now the coin is a $\pi/4$ rotation irrespective of the memory (recall that \mbox{$\theta_C=\pi/4$} also), yielding a balanced walk, regardless of history. But simultaneously the memory register is highly entangled with the position degree of freedom, which, upon being traced out, yields decoherence, which can cause a transition from a quantum to a classical walk. This transition was nicely experimentally demonstrated by Broome \emph{et al.} \cite{bib:Broome10} in the optical context.

If we increase the memory recording strength to \mbox{$\theta_M=\pi/2$}, whilst retaining \mbox{$\theta_B=\pi/4$}, the walker is maximally entangling with the memory, yielding complete decoherence upon tracing out the memory. Thus, in this instance we observe the binomial distribution and linear variance associated with the ideal classical walk (\mbox{Fig.~\ref{fig:composed_symmetric}(e)}).

When \mbox{$\theta_M=0$} there is no memory recording and so no entanglement between the memory and the walker. The memory stays permanently fixed at its initial value and thus the coin is always the \mbox{$\theta_C=\pi/4$} coin, yielding a balanced quantum walk, irrespective of $\theta_B$ (\mbox{Fig.~\ref{fig:composed_symmetric}(c)}). Thus, when \mbox{$\theta_B=\pi/4$}, the memory recording strength can tune us between a perfect classical walk ($\theta_M=\pi/2$) and a perfect quantum walk ($\theta_M=0$).

In the limiting case of maximum coin back-action (\mbox{$\theta_B=\pi/2$}) and maximum memory recording strength (\mbox{$\theta_M=\pi/2$}), the walker resists outward diffusion, yielding a distribution more localised around the origin (\mbox{Fig.~\ref{fig:composed_symmetric}(d)}). This is because as the walker visits sites the respective memory qubit is completely flipped, toggling between marking the site as visited or not visited. When visited, the walker will incur a full bit-flip, sending the walker back whence it came. Otherwise a balanced coin is implemented and the walker spreads equally in both directions. Thus, on alternate steps the walker either completely flips direction or spreads evenly in both directions, resulting in reduced outward diffusion compared to a normal balanced walk.

An interesting feature of \mbox{Fig.~\ref{fig:composed_symmetric}(centre)} is to note that $\sigma^2(\theta_B,\theta_C,\theta_M)$ is not monotonically related to $\theta_B$ and $\theta_M$. For low values of $\theta_B$ we see that $\sigma^2$ increases monotonically with $\theta_M$, whereas for large $\theta_B$ it decreases with $\theta_M$. Similarly, for large $\theta_M$, $\sigma^2$ decreases monotonically with $\theta_B$, whereas for small $\theta_M$ it increases with $\theta_M$. However, the non-monotonic nature of $\sigma^2$ may be due to the small values of $t$ that we are able to simulate. In the case of unsymmetrised input states (Fig.~\ref{fig:3d_asym}) the dynamics are even more complex.

\begin{figure}[!htb]
\includegraphics[width=0.7\columnwidth]{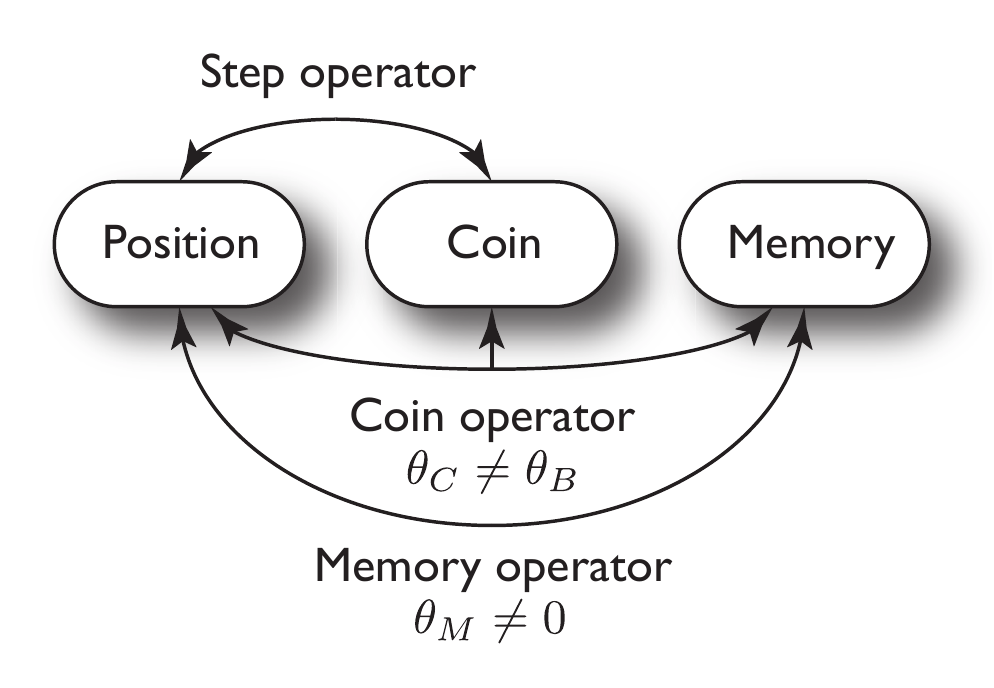}
\caption{Illustration of the couplings between the position, coin and memory subsystems, depending on $\theta_B$, $\theta_C$ and $\theta_M$. The coin operator couples all three subsystems, provided that \mbox{$\theta_C\neq \theta_B$}. The memory operator couples the position and memory subsystems, provided that \mbox{$\theta_M\neq 0$}. The step operator couples the position and coin subsystems, irrespective of the parameters.} \label{fig:coupling}
\end{figure}

Evidently, different parameter regimes for $\theta_B$, $\theta_C$ and $\theta_M$ yield different couplings between the different subsystems when acted upon by the coin, step and memory operators. The couplings are illustrated in Fig.~\ref{fig:coupling}. The coin operator will couple all three subsystems, provided that \mbox{$\theta_C\neq \theta_B$}. The step operator will always couple the coin and position subsystems, irrespective of $\theta_B$, $\theta_C$ and $\theta_M$. And the memory operator will couple the position and memory subsystems provided that $\theta_M\neq 0$.

\section*{Comparison with classical self-avoiding random walks} 
In the classical case, self-avoiding random walks can be constructed in a variety of ways. For example, a classical random walk could be completely self-avoiding, in the sense that it may never return to a visited site. More generally, a metric could be defined, which specifies the probability with which a walker may return to a previously visited site. In Refs.~\cite{bib:Amit83, bib:Byrnes84} such a metric is defined as,
\begin{equation} \label{eq:classical_soft}
q_i = \frac{e^{-g n_i}}{\sum_{j} e^{-g n_j}},
\end{equation}
where $q_i$ is the probability of stepping to site $i$, $n_i$ is the number of times site $i$ has been previously visited, \mbox{$g\geq 0$} is the strength of the self-avoidance (\mbox{$g=0$} is an ordinary classical random walk, \mbox{$g=\infty$} is a completely self-avoiding walk, and \mbox{$0<g<\infty$} is a weakly self-avoiding walk), and $j$ sums over the nearest neighbours of the current vertex. Thus, the probability of returning to a given site is weighted exponentially with the number of times the respective site has been visited. Using this formalism one can tune between a regular non-self-avoiding classical random walk, a weakly self-avoiding random walk, or a completely self-avoiding random walk. A generic characteristic of such a self-avoiding classical random walk is that as $g$ increases the walker is more likely to be found further away from the origin.

The self-avoidance metric given in Eq.~\ref{eq:classical_soft} biases how likely the walker is to revisit a site given its prior visitation history and the strength of the self-avoidance, $g$. However, this bias requires the total number of previous visits to each site. To compare with our quantum self-avoiding walk we restrict to the case where $\theta_M=\pi/2$. Then the memory is marked and then unmarked at subsequent visits, i.e. the bias effectively depends on the number of visits modulo 2. Therefore, we modify Eq.~\ref{eq:classical_soft} to count modulo 2, such that,
\begin{equation} \label{eq:classical_soft_mod}
q_i = \frac{e^{-g (n_i \mathrm{mod}\,2)}}{\sum_{j} e^{-g (n_j \mathrm{mod}\,2)}}.
\end{equation}
Alternately, one could achieve a situation more analogous to Eq.~\ref{eq:classical_soft} by replacing the qubits in the self-avoiding quantum walk with qu-$d$-its and letting the memory recording implement $q_i\to (q_i+1)\,\mathrm{mod}\, d$, thus giving us the ability to record up to $d$ instances of previous visits. However we will not explore this option given the computational intractability.

As discussed earlier, in the quantum case when \mbox{$\theta_M=\pi/2$} we have maximum coupling between the walker and the memory, yielding a decoherence effect. Thus, we expect the comparison between the classical self-avoiding random walk and the quantum self-avoiding walk to be closest in the regime where \mbox{$\theta_M=\pi/2$}.

\begin{figure}[!htb]
\includegraphics[width=0.6\columnwidth]{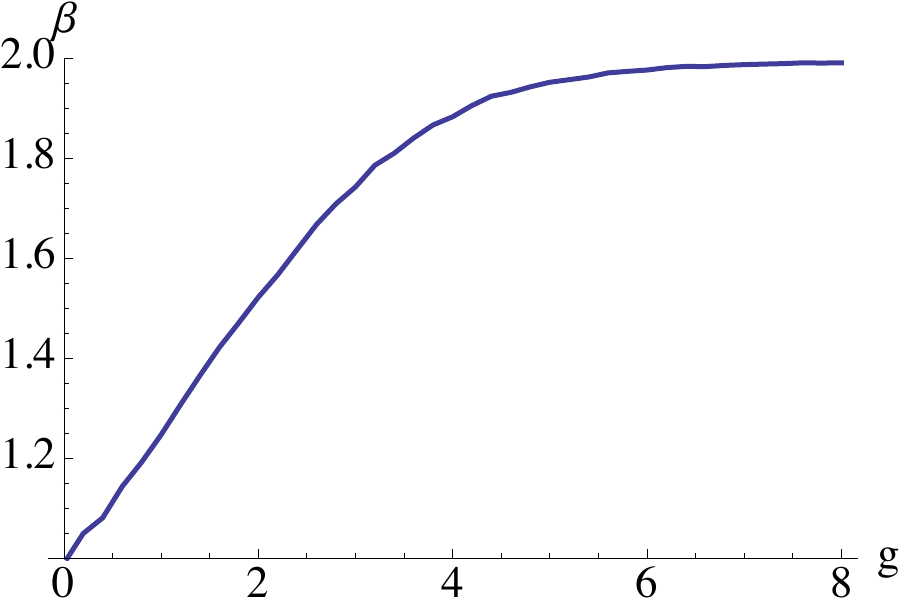} \\
\includegraphics[width=0.6\columnwidth]{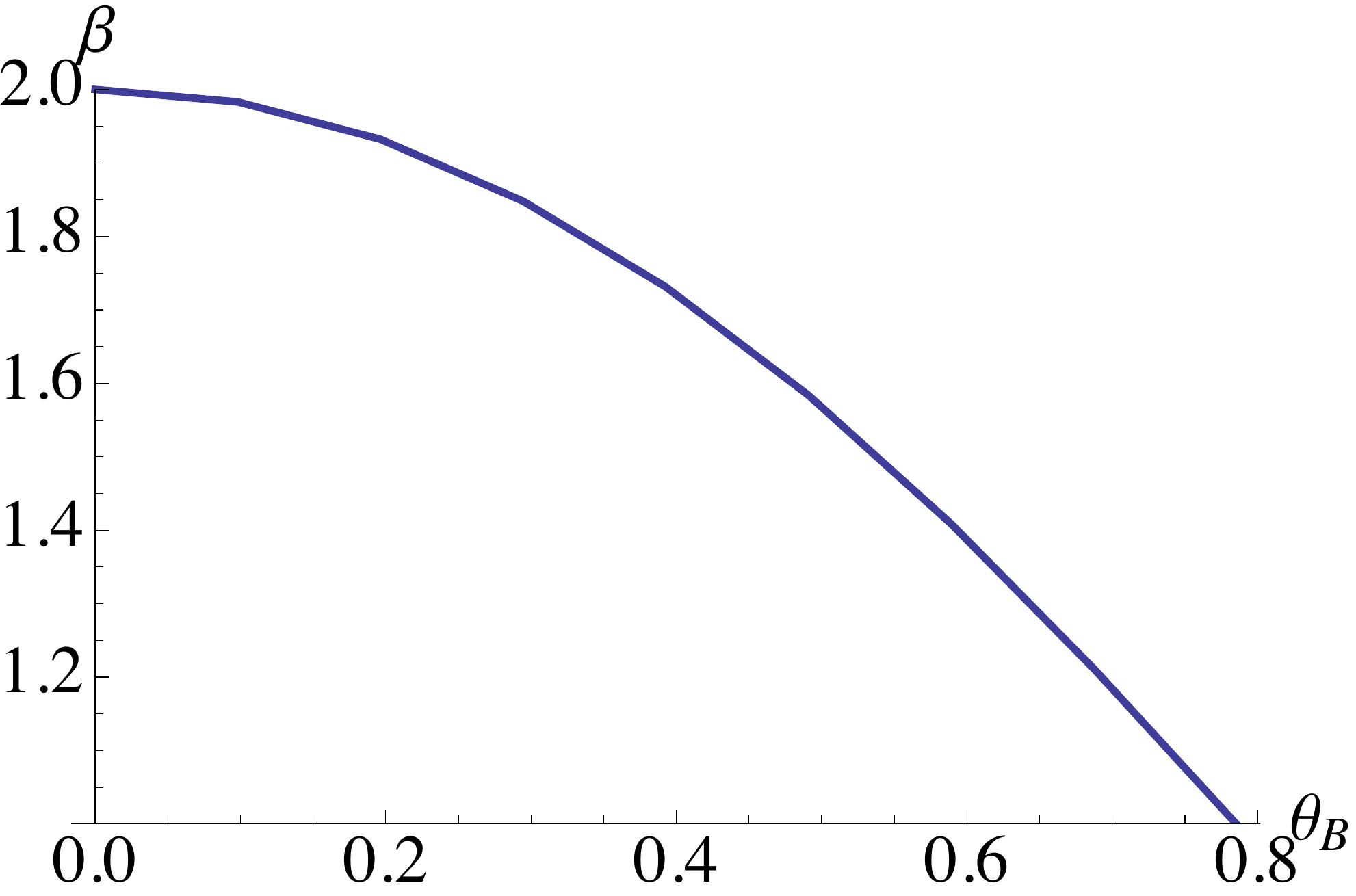}
\caption{Exponent of a \mbox{$\sigma^2(t)=t^\beta$} fit of the variance against time for a self-avoiding walk. \mbox{$\beta=1$} corresponds to linear variance against time, whereas \mbox{$\beta=2$} corresponds to quadratic variance against time (ballistic spread). (top) A classical self-avoiding random walk with self-avoidance function given by Eq.~\ref{eq:classical_soft_mod}, where $g$ is the self-avoidance strength. The fitting is calculated over 200 time steps and 1000 repetitions. When \mbox{$g=0$} we reduce to a normal balanced classical random walk, which exhibits linear spread and thus \mbox{$\beta=1$} (analogous to Fig.~\ref{fig:composed_symmetric}(e)). As the self-avoidance strength increases, we approach a perfectly self-avoiding random walk, exhibiting maximum quadratic dispersion, and \mbox{$\beta=2$} (analogous to Fig.~\ref{fig:composed_symmetric}(a)). (bottom) The same fit for a quantum self-avoiding walk where \mbox{$\theta_C=\pi/4$} and \mbox{$\theta_M=\pi/2$}. This is the classical regime of maximum coupling between the walker and the memory. Now $\theta_B$ takes the role of the self-avoidance strength.} \label{fig:classical_SA_vs_g}
\end{figure}

Recall that in the classical limit \mbox{$\sigma^2(t)\propto t$}, whereas in the quantum limit \mbox{$\sigma^2(t)\propto t^2$}. Thus, to characterise the classical self-avoiding random walk we fit the variance to \mbox{$\sigma^2(t) = t^\beta$}. In Fig.~\ref{fig:classical_SA_vs_g}(top) we plot the exponent $\beta$ against the self-avoidance strength $g$ for a classical self-avoiding random walk. Contrast this with Fig.~\ref{fig:classical_SA_vs_g}(bottom), where we plot the analogous situation for the quantum self-avoiding walk with \mbox{$\theta_M=\pi/2$} while varying $\theta_B$, corresponding to the classical regime of maximum coupling with the memory. Now $\theta_B$ takes the role of the self-avoidance strength. In the limit of \mbox{$\theta_B=\pi/4$} we observe an ideal classical random walk with \mbox{$\beta=1$}, which corresponds to Fig.~\ref{fig:composed_symmetric}(e). In the opposing limit of \mbox{$\theta_B=0$} we observe maximum dispersion and \mbox{$\beta=2$}, corresponding to Fig.~\ref{fig:composed_symmetric}(a). Thus, within this region we observe a correspondence between $g$ (in the classical case) and $\theta_B$ (in the quantum case). Specifically, $g=0$ corresponds to \mbox{$\theta_B=\pi/4$}, and \mbox{$g\gg 0$} corresponds to \mbox{$\theta_B=0$}. In these two limits the self-avoiding classical random walk exhibits dynamics close to the respective self-avoiding quantum walk (Figs.~\ref{fig:composed_symmetric}(e) and (a) respectively). In both cases $\beta$ exhibits a monotonic relationship to the self-avoidance strength, ranging between \mbox{$\beta=1$} (linear spread) and \mbox{$\beta=2$} (ballistic, quadratic spread).


\section*{Conclusion}
We have considered a self-avoiding quantum walk, whereby the walker is complemented by a memory register that records previous position history and the coin operator is a function of the memory.

With appropriate choices of conditional coins and recording strengths, the walk can be made to exhibit zero dispersion or maximal dispersion, which is highly dependent on the symmetrisation of the input state. The walker can be made to reproduce ideal quantum or classical statistics, or to exhibit richer, more diverse diffusive characteristics, which can be manipulated with the strength of the memory recording ($\theta_M$) and back-action ($\theta_B$). In the regime of maximum memory recording strength, we observe a close correspondence between the quantum and classical self-avoiding walks, where the coin back-action takes the role of the self-avoidance strength.

The rich diffusive characteristics of the quantum self-avoiding walk may be applicable to modelling a broader range of diffusive phenomenon than ordinary quantum walks, and may lend themselves to novel quantum information processing applications such as applications in quantum simulation.


\begin{acknowledgments}
This research was conducted by the Australian Research Council Centre of Excellence for Engineered Quantum Systems (Project number CE110001013).
\end{acknowledgments}

\section*{Additional Information}
The authors declare no competing financial interests.


\section*{Author Contributions}
The central idea for this work was suggested by JT and furthered developed in discussions between JT and PR and EC. EC and PR primarily did the numerics and all the authors wrote and reviewed the paper.

\end{document}